\documentclass[namedreferences]{kluwer}
\usepackage[]{graphicx}
\begin{document}
\begin{article}
\begin{opening}
\title{Low-Mass Cluster Galaxies: A Cornerstone of Galaxy Evolution} 
           

\author{\surname{Christopher J. Conselice}\email{cc@astro.caltech.edu}}
\institute{California Institute of Technology, Pasadena CA}         
                      




\runningtitle{Low-Mass Cluster Galaxies}
\runningauthor{Christopher J. Conselice}



\begin{abstract} 
Low-mass cluster galaxies are the most common galaxy type in the universe
and are at a cornerstone of our understanding of galaxy formation, cluster 
luminosity functions, dark matter and the formation of large scale
structure.  I describe in this summary recent observational results 
concerning the properties and likely origins of low-mass galaxies in clusters
and the implications of these findings in broader galaxy formation issues.

\end{abstract}




\end{opening}
\section{Low-Mass Cluster Galaxies}

Although they are the faintest and lowest mass galaxies in the universe, 
low-mass cluster galaxies (LMCGs), especially dwarf ellipticals, hold
clues for the ultimate
understanding of galaxy formation, dark matter and structure formation.  
The reason for this is quite simple: low-mass galaxies, and particularly 
low-mass galaxies in clusters (Conselice et al. 2001) are the most common 
galaxies in the nearby universe (Ferguson \& Binggeli 1994).  Any ultimate 
galaxy evolution/formation scenario must be able
to predict and accurately describe the properties of these objects.  
In galaxy formation models, such as hierarchical assembly (e.g., 
Cole et al. 2000), massive 
dark halos form by the mergers of lower mass ones early in the universe.  By
understanding these LMCGs, we are potentially studying the very first
galaxies to form.   On the other hand, observations
reveal that no low-mass galaxies formed all of their stars early in the 
universe at $z > 7$,
with considerable evidence for star formation occurring in the last
few Gyrs (e.g., Grebel 1997; Conselice et al. 2003).

While low-mass galaxies are traditionally studied in low density
environments, such as in the Local Group, it is now clear that a large
population of these low-mass galaxies exist in clusters,
whose nature is only recently becoming clear (Conselice
et al. 2001; 2003).  A comparison with the Local Group demonstrates that the 
ratio of low-mass to large mass galaxies in clusters is roughly five to ten 
times higher than in low density environments.  This over density of LMCGs, 
and the fact that some 
Local Group dwarf spheroidals (Klyena et al. 2002) have 
large dark matter halos, hints that potentially a large amount of mass in
clusters is associated with low-mass galaxies.  New observational
results also suggest that early-type LMCGs are not a homogeneous population, 
but consist of at least two distinct types, that are morphologically similar, 
but with different physical properties.

\section{New Observational Results}

There are several observations, listed below, that suggest low-mass cluster
galaxies have unique dynamical, kinematic and stellar population properties
that differ from properties of Local Group low-mass galaxies (see e.g., 
Conselice, Gallagher
\& Wyse 2001, 2002, 2003; Rakos et al. 2001; Pedraz et al. 2002).   

\begin{enumerate}

\item {\bf Spatial Position:} While Local Group dwarf galaxies, particularly
dwarf ellipticals, are strongly clustered around the giant galaxies
in the Local Group (van den Bergh 2000), the opposite is found for low-mass 
galaxies in clusters, where most are neither clustered around, nor 
distributed globally similar to, the giant elliptical galaxies 
(Conselice et al. 2001).

\begin{figure}[H]
\tabcapfont
\centerline{%
\begin{tabular}{c@{\hspace{6pc}}c}
\includegraphics[width=2.7in]{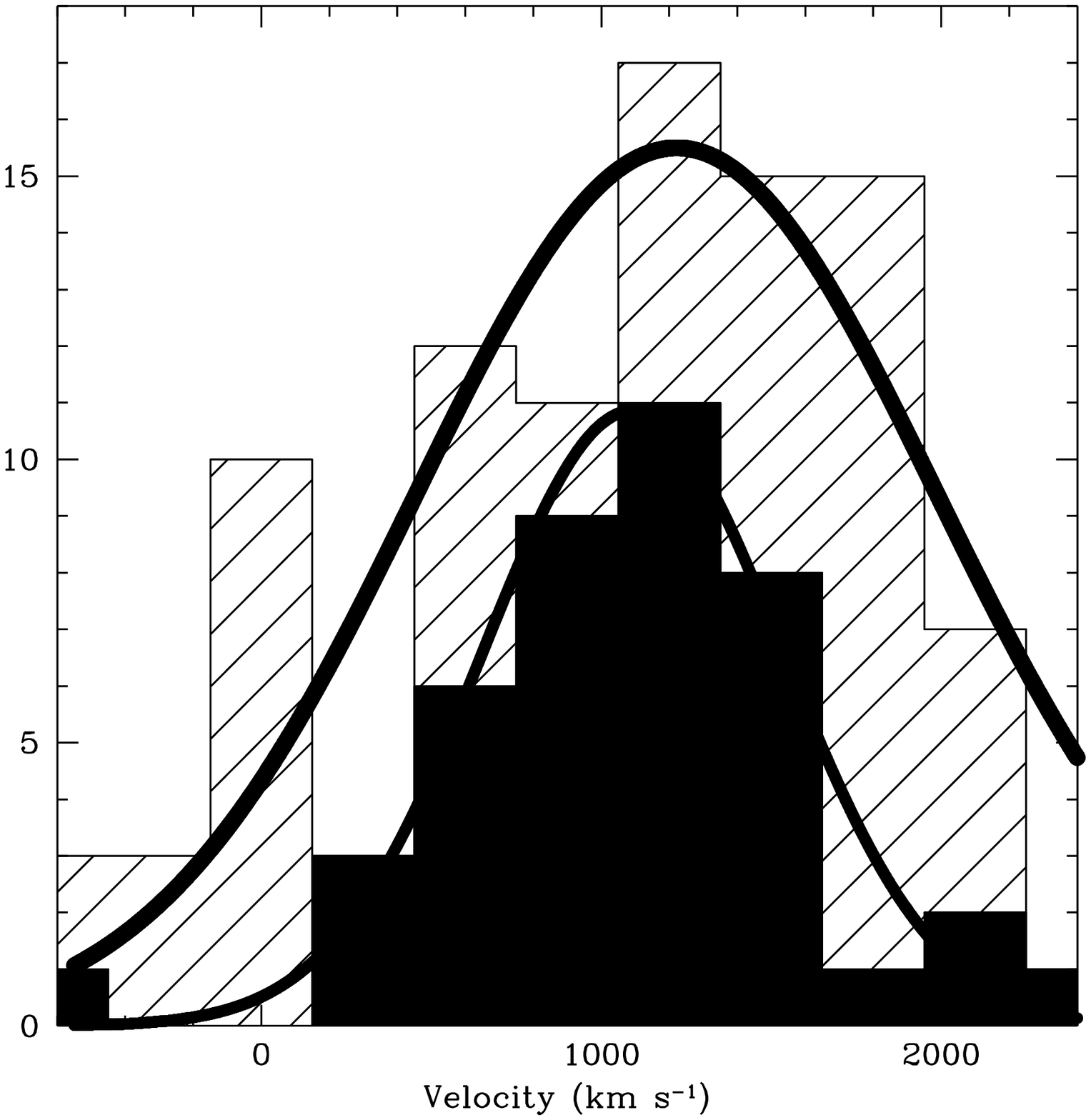} &
\hspace{-1.5cm}
\includegraphics[width=3.0in]{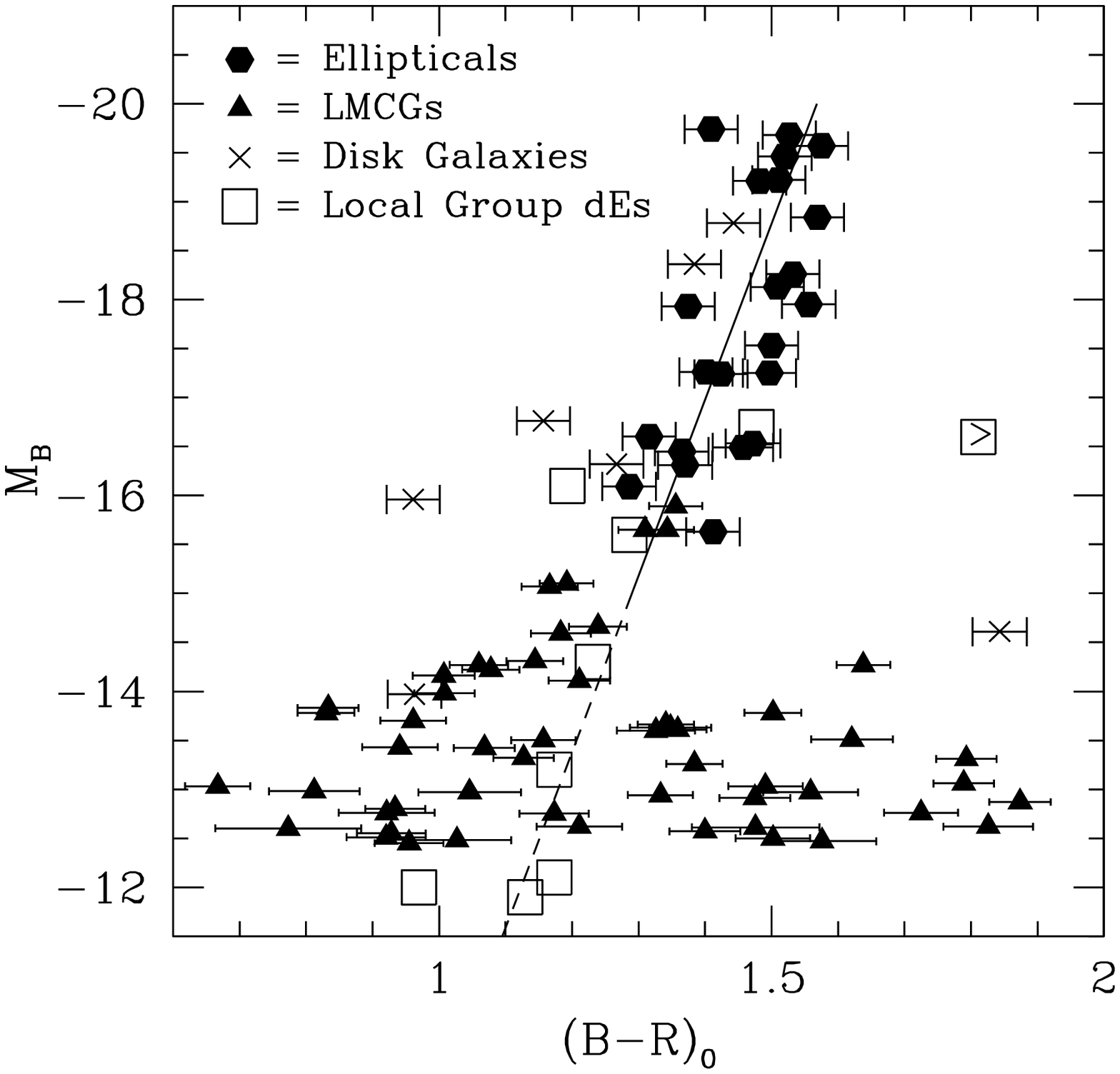} \\
\end{tabular}}
\caption{(a) Velocity histograms for giant ellipticals (solid) and dwarf 
ellipticals (shaded)
in the Virgo cluster (Conselice et al. 2001) (b) Color magnitude diagram for galaxies in the Perseus cluster, 
demonstrating the large color scatter for systems with M$_{\rm B} > -15$.
The solid boxes are where Local Group dEs/dSphs would fit on this plot.}
\end{figure}

\item {\bf Radial Velocities:} The radial velocities of low-mass
cluster galaxies, including S0s, spirals, dwarf irregulars and dwarf
ellipticals are more widely distributed than the ellipticals (see
Figure~1a).  For example, Virgo cluster elliptical
galaxies have a narrow Gaussian velocity distribution, with $\sigma = 462$ 
km s$^{-1}$, concentrated at the mean 
radial velocity of the cluster.  The other populations, including
the over 100 classified dwarf ellipticals in Virgo with radial
velocities, have much broader, and non-Gaussian, velocity distributions 
($\sigma \sim 700$ km s$^{-1}$), all with velocity dispersion ratios with 
the ellipticals consistent with their being accreted (e.g., 
Conselice et al. 2001).

\item {\bf Stellar Populations:} Faint LMCGs, with 
M$_{\rm B} > -15$, have a large color scatter at a given magnitude 
(e.g., Rakos et al. 2001; Conselice et al. 2003) produced by galaxies
that are both bluer and redder than the extrapolated color-magnitude
relationship, as defined by the giant elliptical galaxies (Figure~1b).  
This is found in
several nearby clusters, including Fornax, Coma and Perseus, and can be 
explained by the stellar populations in LMCGs having a mixture of ages and 
metallicities (e.g., Rakos et al. 2001; Conselice et al. 2003).  Stromgren
and broad-band photometry reveals that the red LMCGs are metal enriched
systems.  These red LMCGs 
steepen the luminosity function (LF) of clusters, and are responsible for 
differences in faint end LF slopes seen in clusters and in 
the field (Conselice 2002).

\item {\bf Internal Kinematics:} Some LMCGs show evidence for rotation when
observed out to at least one scale radii (e.g.,
Pedraz et al. 2002).   Rotation is however not present in Local
Group dEs, such as NGC 205 and NGC 185 (e.g., van den Bergh 2000). 

\end{enumerate}

\section{LMCG Origins}

Based on the observational results presented above it appears that some
LMCGs are fundamentally different than low-mass galaxies in groups, although 
bright LMCGs have similar photometric properties to Local Group dEs  
(e.g., Conselice et al. 2003). 

Several possible physical mechanisms can potentially explain the origin of 
LMCG populations.  In the simple collapse + 
feedback scenario (Dekel \& Silk 1986), LMCGs are formed when gas collapses 
and forms stars. These stars produce winds that expels gas from these systems,
halting any future star formation.
In this formation scenario LMCGs formed before the cluster
ellipticals, or at least formed within groups that later merged to form 
clusters. Faint LMCGs 
however, cannot all be born in groups which later accreted into 
clusters along with the massive galaxies, due to the high LMCG to giant 
galaxy ratio found in clusters (Conselice et al. 2001, 2003).
The above evidence suggests that simple low-mass galaxy formation scenarios
can be safely ruled out for some LMCGs.

One alternative idea is that present day LMCGs
formed after the cluster itself was in place by collapsing out
of enriched intracluster gas.  Another is that the intracluster medium (ICM) 
is able to retain enriched gas that in the Dekel and Silk
(1986) paradigm would be ejected by feedback, but remains due to the 
confinement pressure of the ICM (Babul \& Rees 1992).  This scenario would 
explain the higher metallicities of some of the fainter LMCGs.

An alternative scenario is that LMCGs form in the cluster through
a tidal origin.   Two main possibilities for this are tidal dwarfs (Duc
\& Mirabel 1994), and as the remnants of 
stripped disks or dwarf irregulars (Conselice et al. 2003).
The velocity and spatial distributions of LMCGs suggest that they must
have been accreted into the cluster during the last few Gyrs 
(Conselice et al. 2001).  This, combined
with the high metallicities of these LMCGs, and the fact that their
stellar populations are fundamentally different than field dwarfs (e.g.,
Conselice 2003; Figure 1b) suggests that the cluster environment has
morphologically transformed accreted galaxies into LMCGs.  This is
consistent with the internal rotation found for some LMCGs (Moore
et al. 1998). Ongoing and future observations of the HI, dynamical and dark
properties of LMCGs will soon allow for a more complete 
observational description of these objects.

\end{article}

\vspace{-0.7cm}
\begin{thebibliography}{}
\bibitem[\protect\citeauthoryear{}{}]{} Babul, A., \& Rees, M.J. 1992, MNRAS, 255, 346 
\bibitem[\protect\citeauthoryear{}{}]{} Conselice, C.J., Gallagher, J.S., \& Wyse, R.F.G. 2001, ApJ, 559, 791 
\bibitem[\protect\citeauthoryear{}{}]{} Conselice, C.J., Gallagher, J.S., \& Wyse, R.F.G. 2002, AJ, 123, 2246 
\bibitem[\protect\citeauthoryear{}{}]{} Conselice, C.J., Gallagher, J.S., \& Wyse, R.F.G. 2003, in press, astro-ph/0210080 
\bibitem[\protect\citeauthoryear{}{}]{} Conselice, C.J. 2002, ApJ, 573, 5L 
\bibitem[\protect\citeauthoryear{}{}]{} Cole, S. et al. 2000, MNRAS, 319, 168 
\bibitem[\protect\citeauthoryear{}{}]{} Dekel, A., \& Silk, J. 1986, ApJ, 303, 39
\bibitem[\protect\citeauthoryear{}{}]{} Duc, P.-A., \& Mirabel, I.F. 1994, A\&A, 289, 83 
\bibitem[\protect\citeauthoryear{}{}]{} Ferguson, H.C., \& Binggeli, B. 1994, A\&ARv, 6, 67 
\bibitem[\protect\citeauthoryear{}{}]{} Grebel, E.K. 1997, RvMA, 10, 29 
\bibitem[\protect\citeauthoryear{}{}]{} Kleyna, J., et al. 2002, MNRAS, 330, 792
\bibitem[\protect\citeauthoryear{}{}]{} Moore, B. et al. 1998, ApJ, 495, 139 
\bibitem[\protect\citeauthoryear{}{}]{} Pedraz, S. et al. 2002, MNRAS, 332, 59L 
\bibitem[\protect\citeauthoryear{}{}]{} Rakos, K. et al. 2001, AJ, 121, 1974 
\bibitem[\protect\citeauthoryear{}{}]{} van den Bergh, S. 2000, The Galaxies of the Local Group, Cambridge Univ. Press 

\end{thebibliography}
\end{document}